\documentclass[twocolumn,preprintnumbers,amsmath,amssymb,superscriptaddress,longbibliography,nofootinbib,prb]{revtex4-1}

\usepackage{graphicx}
\usepackage{bm}
\usepackage{dsfont}
\usepackage[usenames,dvipsnames]{xcolor}
\usepackage{pstricks}
\usepackage[tight]{subfigure}
\usepackage{verbatim}
\usepackage{units}
\usepackage{multirow}
\usepackage{enumitem}
\usepackage{mathrsfs}
\usepackage{leftidx}
\usepackage{xspace}

\usepackage{amsfonts}

\usepackage{hyperref}
\hypersetup{colorlinks=true,linktoc=all,linkcolor=blue,breaklinks=true,citecolor=blue,urlcolor=blue}

\usepackage[english]{babel}
\usepackage[normalem]{ulem}

\AtBeginDocument{%
    \newwrite\bibnotes
    \def\bibnotesext{Notes.bib}
    \immediate\openout\bibnotes=\jobname\bibnotesext
    \immediate\write\bibnotes{@CONTROL{REVTEX41Control}}
    \immediate\write\bibnotes{@CONTROL{%
    apsrev41Control,author="08",editor="1",pages="1",title="0",year="1"}}
     \if@filesw
     \immediate\write\@auxout{\string\citation{apsrev41Control}}%
    \fi
}%

\begin{document}
	
	
	\title{Mesoscopic effects in the heat conductance of superconducting--normal--superconducting
and normal--superconducting junctions}
	
	\author{Fatemeh Hajiloo}
	\affiliation{Department of Microtechnology and Nanoscience (MC2), Chalmers University of Technology, S-412 96 G\"oteborg, Sweden}

	\author{Fabian Hassler}
	\address{JARA-Institute for Quantum Information, RWTH Aachen University, D-52074 Aachen, Germany}
		
	\author{ Janine Splettstoesser}
	\affiliation{Department of Microtechnology and Nanoscience (MC2), Chalmers University of Technology, S-412 96 G\"oteborg, Sweden}
	
	\date{\today}
	
	\begin{abstract}
	We study the heat conductance of hybrid superconducting junctions. Our analysis involves single-channel junctions with arbitrary transmission as well as diffusive connectors and shows the influence of the superconducting gaps and phases of the contacts on the heat conductance. If the junction is \textit{diffusive}, these effects are completely quenched on average, however, we find that their influence persists in weak-localization corrections and conductance fluctuations.  While these statistical properties strongly deviate from the well-known analogues for the charge conductance, we demonstrate that the heat conductance fluctuations maintain a close to universal behavior. We find a generalized Wiedemann-Franz law for Josephson junctions with equal gaps and vanishing phase difference. 
	\end{abstract}

	\maketitle

\section{Introduction}\label{sec_intro}

 The conductance of heat in hybrid superconducting junctions plays a crucial role in coherent caloritronics and for quasiparticle cooling. In coherent caloritronics~\cite{Giazotto2012Dec,Fornieri2017Oct}, the phase-difference across Josephson junctions consisting of superconductors separated by an insulating  (SIS) or a normalconducting layer (SNS) is used to tune heat transport coherently. Coherent effects in the heat transport occur also in other types of hybrid superconducting devices, such as Andreev interferometers~\cite{Bezuglyi2003Sep,Chandrasekhar2009Jul}. For quasiparticle cooling~\cite{Muhonen2012Mar,Courtois2014Jun} the difference of gaps across a junction ---such as in NS junctions, where one of the gaps completely vanishes--- are used as an energy filter. With devices getting ever more complex, it is of crucial importance to understand the impact of the junction properties on the heat transport characteristics.

In this paper, we analyze the heat conductance of SNS junctions with an arbitrary phase difference and arbitrary magnitude of the two gaps of the two superconductors (including the NS limit). We treat both the case of single-channel junctions as well as diffusive junctions, in which the transmission probabilities of the many transport channels are statistically distributed. We analyze the average of the heat conductance, its weak-localization correction, as well as heat conductance fluctuations, by combining a scattering matrix approach to heat transport~\cite{Butcher1990Jun} for superconducting junctions~\cite{Beenakker1991Dec} with previously obtained results from random matrix theory~\cite{Beenakker1997Jul}.

The statistics of normalconducting diffusive junctions~\cite{Beenakker1997Jul}, including the famous weak-localization effects~\cite{Altshuler1980Dec,Cahay1988Jun} and universal conductance fluctuations~\cite{Lee1985Oct,Altshuler1985Jun,Imry1986Mar,Altshuler1991Jul} have been analyzed more than two decades ago and have been among the most fundamental properties of mesoscopic devices. Also, modifications in hybrid devices have been studied~\cite{Hecker1997Aug,Heikkila1999Oct}.
However, these statistical properties of the \textit{heat} conductance of diffusive junctions have to our knowledge not been addressed beyond the statistical average~\cite{Bezuglyi2003Sep,Yokoyama2005Dec,Vasenko2010Mar,Virtanen2015Feb,Pershoguba2019Apr}. Indeed, in fully normalconducting junctions with energy-independent transmission probabilities, heat and charge conductances ---both being linearly dependent on the transmission probabilities of the junction channels--- are up to a different conductance quantum the same due to the Wiedemann-Franz law. This is however completely different for Josephson junctions where only  the linear heat conductance assumes a finite value~\cite{Zhao2003Aug,Zhao2004Apr} in contrast to the charge conductance that is undefined due to the presence of the supercurrent. Furthermore, the complex energy-dependence of the transmission probabilities of SNS and NS junctions leads to important differences in the statistics of the heat conductance with respect to the charge conductance.

\begin{figure}[b]
	\includegraphics[width=0.9\linewidth]{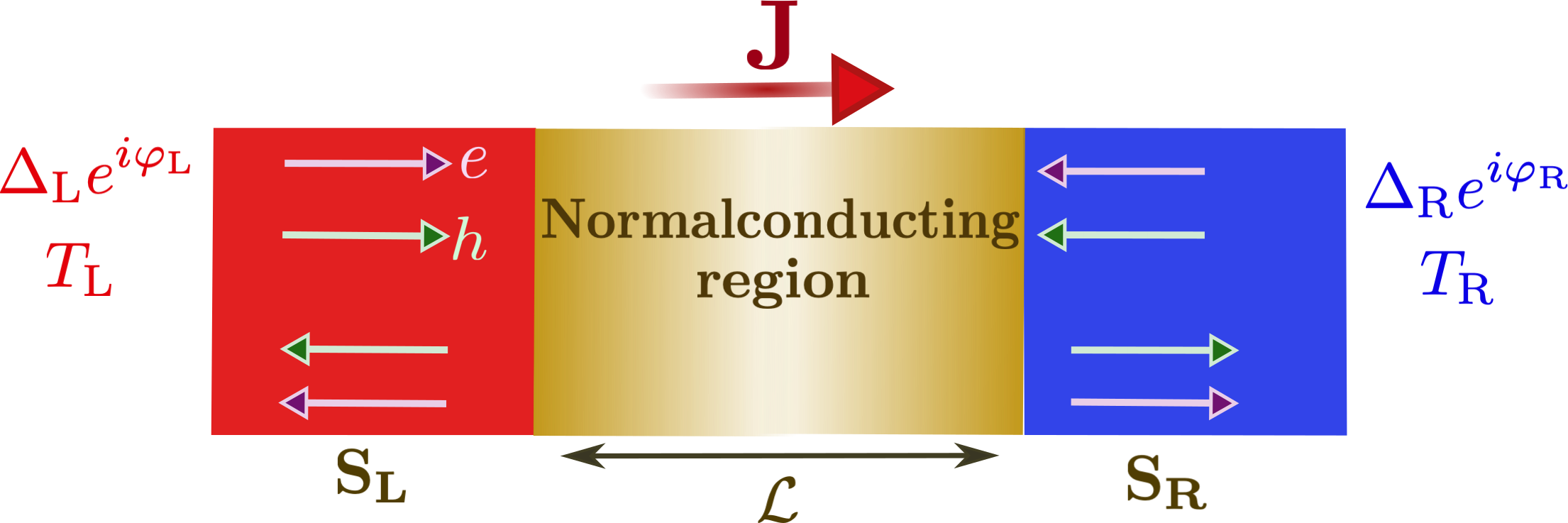}
\caption{Sketch of an SNS junction across which a heat current $J$ flows. The contacts are characterized by different temperatures, $T_\text{L}$ and $T_\text{R}$, superconducting gaps, $\Delta_\text{L}$ and $\Delta_\text{R}$ ---one of which can even be suppressed to 0, and phases, $\varphi_\text{L}$ and $\varphi_\text{R}$. The normal part of the junction has length $\mathcal{L}$ and supports $\mathcal{N}$ scattering channels. In this paper, we treat the case of $\mathcal{N}=1$ and the case of a diffusive region with a non-specified number of channels.\label{fig_model} }
\end{figure}

Here, we present results for the heat conductance of SNS or NS junctions, see Fig.~\ref{fig_model}, which can exceed the heat conductance of normal conducting junctions by a considerable amount, due to the phase-difference~\cite{Zhao2003Aug,Zhao2004Apr} or differences in superconducting gaps. These effects are completely quenched in the average heat conductance of a diffusive junction~\cite{Virtanen2015Feb}. The weak-localization corrections as well as the heat conductance fluctuations, both relying on quantum interference effects, however maintain a dependence on the phase difference and variations in the gap magnitude of the contacts. Importantly, despite these nontrivial dependencies, the heat conductance fluctuations change by less than an order of magnitude under parameter variation, hence remaining close to universal. Furthermore, for junctions with equal gaps and vanishing phase-difference we identify a generalized Wiedemann-Franz law.

SNS junctions with different distributions of transmission probabilities have been realized with distinct techniques and materials. Metallic systems can contain diffusive N-junctions of different length, see e.g. Ref.~\onlinecite{Dubos2001Jan}, in break junctions the transmission of the few contributing channels depends on specific electronic configurations of the junction~\cite{Bretheau2013Jul}, and superonductor-semiconductor junctions~\cite{Krogstrup2015Jan,Kjaergaard2016Sep,Larsen2015Sep} can be realized with  tunable channel number  with possibly statistically distributed transmissions~\cite{vanWoerkom2017Jun}. This variety of SNS junction realizations makes the experimental verification of the results predicted here likely.

\section{Model and approach}\label{sec_model}

 We study an SNS junction as sketched in Fig.~\ref{fig_model}. The contacts are assumed to be standard BCS superconductors with a temperature- and phase-dependent order parameter $\Delta_\alpha=|\Delta_\alpha(T)|\text{exp}(i\varphi_\alpha)$. The temperature-dependent absolute value of the gap $|\Delta_\alpha(T)|$ is obtained self-consistently with $\Delta_{0,\alpha}\equiv|\Delta_\alpha(0)|\simeq1.76 k_\text{B}T_{\text{crit},\alpha}$. We take the gap function to be space-independent within each segment of the junction, which is justified when the diffusion constant in the normalconducting link is larger than the diffusion constant in the superconductor~\cite{Likharev1979Jan,Beenakker1992Jan}.
The two superconducting contacts $\alpha=$L,R can have different gaps $\Delta_{0,\text{L}},\Delta_{0,\text{R}}$ (and hence different critical temperatures $T_{\text{crit},\text{L}},T_{\text{crit},\text{R}}$) and they have a possible phase difference, $\varphi=\varphi_\text{L}-\varphi_\text{R}$.
Different temperatures, $T_\text{L,R}=T\pm\frac12\delta T$, lead to a heat flow between the contacts. Importantly, all of these quantities characterizing the contacts are separately tunable. We are not interested in the effect of a  bias voltage and therefore, from here on, take the common electrochemical potentials as the zero of energy $\mu_\text{L}=\mu_\text{R}\equiv0$. 
The normal conducting junction has length $\mathcal{L}$; the finite width of the junction leads to a number $\mathcal{N}$ of transport channels.  In the diffusive regime, the length of the junction is assumed to be much larger than the mean free path, $\ell$, but much shorter than the localization length, $\mathcal{N}\ell$, and the dephasing length. The assumption of phase-coherent particle transport across the normal region is reasonable for the low-temperature limit (temperatures not exceeding the critical temperature of s-wave superconductors such as aluminum) considered here. 

Heat transport across this junction is carried by electron- and hole-like quasiparticles ($e,h$), which can be transmitted across the SN interfaces by normal transmission and Andreev reflection.  We here use the Andreev approximation, meaning that both electrons and holes travel approximately at the Fermi velocity~\cite{deGennes1999}. In order to describe heat transport, we use a scattering matrix approach. The scattering matrix $\mathcal{S}_\text{SNS}$ relates the fluxes carried by quasiparticles of different incoming and outgoing channels to each other. We assume that Andreev reflection at the NS interfaces does not mix channels and diagonalize the channel-mixing normal part of the scattering matrix using a polar decomposition. See, e.g., 
Ref.~\onlinecite{Beenakker1991Dec} for the construction and diagonalization of the $\mathcal{N}$-channel scattering matrix.
We find that the transmission probabilities, $\mathcal{D}_n^{ij}$ for $i,j=e,h$, characterizing $\mathcal{S}_\text{SNS}$,  each depend on a single transmission eigenvalue $D_n$ of the normalconducting junction only. 
The transmission probability $\mathcal{D}_n^e$ of channel $n$ is the sum of the transmission probabilities from electron- and hole-like quasiparticles into an electron-like quasiparticle channel, $\mathcal{D}_n^e=\mathcal{D}_n^{ee}+\mathcal{D}_n^{eh}$. 
We find the transmission probability for arbitrary gaps $\Delta_\text{L}$ and $\Delta_\text{R}$  to be given by~\cite{Zhao2004Apr,Virtanen2015Feb}
\begin{align}
\label{eq_fullTransmission}
&\mathcal{D}_n^e(E) = \\
&  2D_n \xi_{\text{L}}\xi_{\text{R}}\frac{D_n \xi_{\text{L}}\xi_{\text{R}}+(2-D_n)(E^{2}-|\Delta_{\text{L}}\Delta_{\text{R}}|\cos\varphi)}{((2-D_n)\xi_{\text{L}}\xi_{\text{R}}+D_n(E^{2}-|\Delta_{\text{L}}\Delta_{\text{R}}|\cos\varphi))^{2}},\nonumber
\end{align} 
where $\xi_{\alpha}=\sqrt{E^{2}-|\Delta_{\alpha}|^{2}}$ is the quasiparticle energy in contact $\alpha$. This transmission probability is generally finite for energies $E>|\Delta|$, with $|\Delta|:=\text{max}(|\Delta_\alpha(T_\alpha=T)|)$, and zero otherwise.
As a result, the heat current can be written as a sum over $\mathcal{N}$ transport eigenchannels
\begin{equation}\label{eq_heatcurrent}
J=\frac{2}{h}\sum_{n=1}^\mathcal{N}\int_{|\Delta|}^\infty dE \,E \ \mathcal{D}_n^e(E)\left(f_\text{L}(E)-f_\text{R}(E)\right)\ .
\end{equation}
Here, Fermi functions $f_\alpha(E)=\left[1+\exp(E/k_\text{B}T_\alpha)\right]^{-1}$ determine the quasiparticle occupation in contact $\alpha$. 
The factor 2 in Eq.~(\ref{eq_heatcurrent}) is due to  spin degeneracy~\cite{Chtchelkatchev2003Jun}. 

\section{Heat current and heat conductance}\label{sec_conductance} 

%
From Eqs.~(\ref{eq_fullTransmission}) and (\ref{eq_heatcurrent}), we can straightforwardly derive the heat conductance, $\kappa_\text{SNS}=\partial J/\partial\delta T|_{\delta T=0}$, that is of interest for small temperature gradients $\delta T$,
\begin{equation}
\label{eq_full_cond}
\kappa_\text{SNS}=\frac{1}{2h}\sum_{n=1}^\mathcal{N}\frac{1}{k_\text{B}T^{2}}\int_{|\Delta|}^{\infty}dE \frac{E^{2}}{\cosh^{2}(\frac{E}{2k_\text{B}T})}\left.\mathcal{D}_n^e(E)\right|_{\delta T=0}\ .
\end{equation}
For superconducting contacts with $\Delta_{0,\text{L}}=\Delta_{0,\text{R}}$, we recover previously obtained results in the single-~\cite{Spilla2015Jun} and multi-channel regimes~\cite{Zhao2003Aug,Zhao2004Apr}. Note that results obtained in the tunneling limit~\cite{Maki1965Dec,Guttman1997May,Golubev2013Mar,Spilla2014Apr}, which can not account for the creation of Andreev bound states, are in equal-gap junctions only valid for heat currents at relatively large temperature gradients and the heat \textit{conductance} is hence in general not straightforwardly obtained from this.

\begin{figure}[t]
	\includegraphics[scale=1]{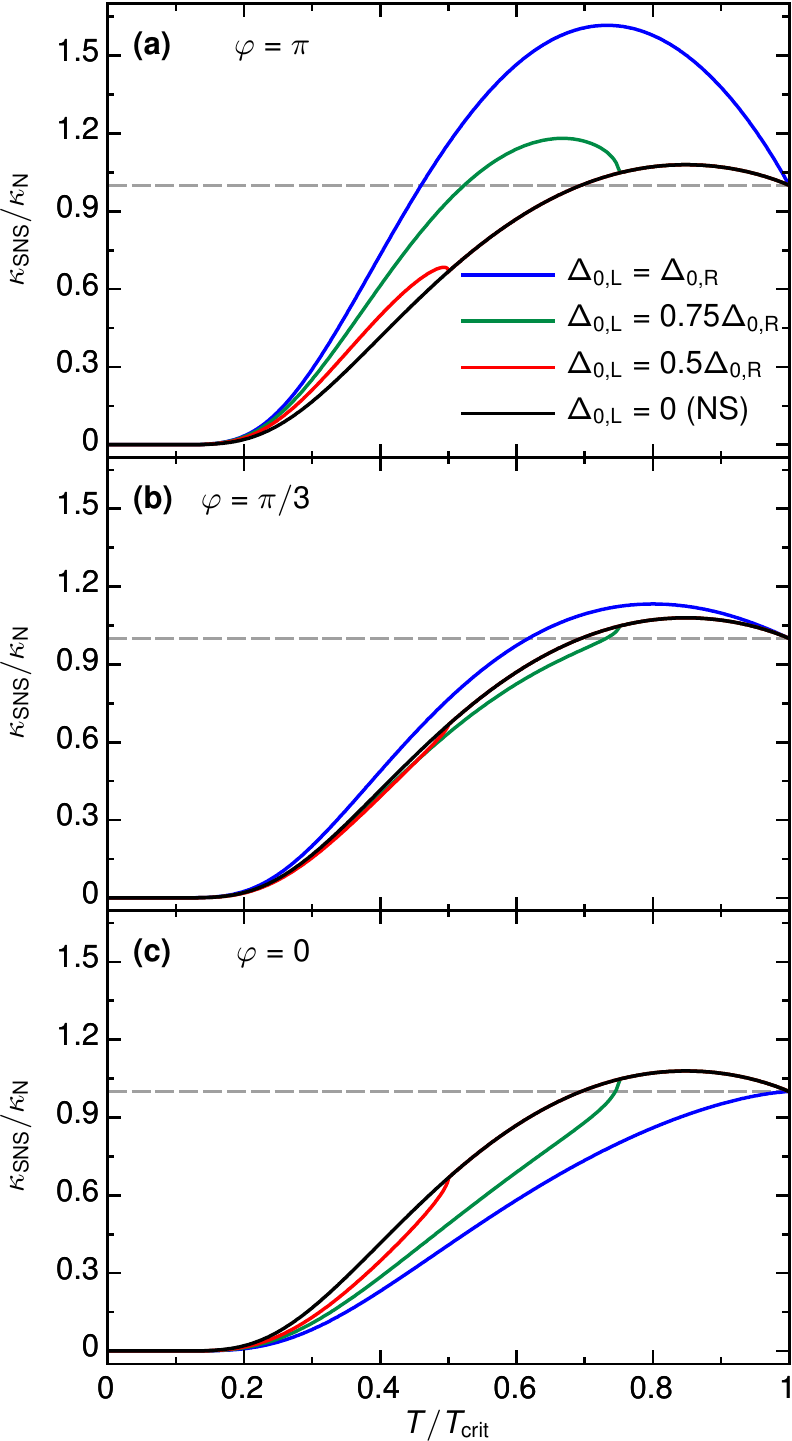}
	\caption{
	Heat conductance as function of $T/T_\text{crit}$, with the critical temperature of the larger gap $T_\text{crit}\equiv T_\text{R,crit}$. Panels (a) to (c) show the single-channel case with  $D_n=D=0.1$ for different values of $\varphi$ and $\Delta_{0,\text{L}}/\Delta_{0,\text{R}}$. The grey dashed line at $\kappa_\text{SNS}/\kappa_\text{N}=1$ serves as a guide for the eye.}
	\label{fig:singlechannel}
\end{figure}

The heat conductance, Eq.~(\ref{eq_full_cond}) with $\mathcal{D}_n^e(E)$ from Eq.~(\ref{eq_fullTransmission}), is shown in Fig.~\ref{fig:singlechannel}, for a single-channel junction with transmission $D_1=D=0.1$ for different phase-differences $\varphi$ and different gap ratios $\Delta_{0,\text{L}}/\Delta_{0,\text{R}}$. We choose a normalization of $\kappa_\text{SNS}$  with respect to the value of a fully normalconducting  (single-channel) device $\kappa_\text{N}=\kappa_0\sum_{n=0}^\mathcal{N}D_n\rightarrow\kappa_0D$. Here,  $\kappa_0=\frac{\pi^2 k_\text{B}^2}{3h}T$ is the (temperature-dependent) heat conductance quantum~\cite{Pendry1983Jul}, recently measured in electronic systems~\cite{Jezouin2013Nov}.  This choice of normalization highlights that, depending on phase and transmission, the heat conductance $\kappa_\text{SNS}$ can exceed the normalconducting case, as previously found for $\Delta_{0,\text{L}}=\Delta_{0,\text{R}}$ in Ref.~\onlinecite{Zhao2004Apr}. 
Here, we find that different gaps have an important influence on this predicted behavior. Independently of the phase difference, $\kappa_\text{SNS}$ can exceed $\kappa_\text{N}$ as soon as one gap is smaller than the other; this is in particular true for the NS-case, shown as a black line, where one of the gaps completely vanishes. This limit is also reached for $\Delta_{0,\text{L}}\neq0$, when the temperature exceeds the critical temperature $T_\text{L,crit}\neq T_\text{R,crit}$ (visible as a kink, before the lines start to overlap with the black line of the NS result in panels (a) to (c) in Fig.~\ref{fig:singlechannel}).  For the low-transmission limit ($D=0.1$) considered here, we conclude the following: (i) equal gaps are favorable for the heat conductance at large phase differences and (ii) different gaps (in particular the NS limit) maximally increase the heat conductance, if the phase differences are  small or vanishing. The reason for the increase of the heat conductance $\kappa_\text{SNS}$ with respect to $\kappa_\text{N}$ is an energy-dependent modulation of the transmission probability $\mathcal{D}_n^e(E)$, which can result into $\mathcal{D}_n^e(E)>D$ in the vicinity of the gap $|\Delta|$, and hence at relatively large energies. For the equal-gap case, this modulation was attributed to a phase-dependent Andreev bound state~\cite{Zhao2004Apr}. In the opposite limit of one vanishing gap (NS case) no bound states arise and Andreev reflection above the gap is responsible for modulations of the transmission probability~\cite{Blonder1982Apr}. Note that for larger transmissions $D>0.5$, the modulation of the heat conductance $\kappa_\text{SNS}$ always leads to a reduction with respect to $\kappa_\text{N}$.

\begin{figure}[t]
	\includegraphics[scale=1]{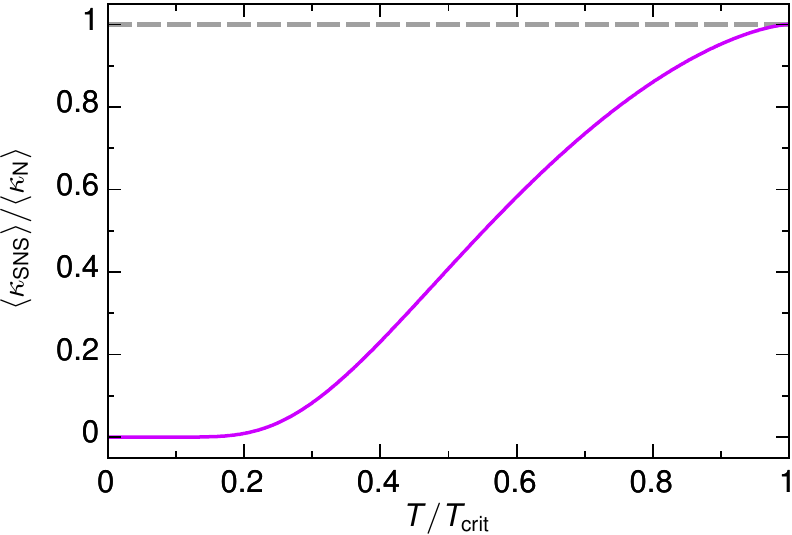}
	\caption{
 The average heat conductance $\langle\kappa_\text{SNS}\rangle/\langle\kappa_\text{N}\rangle$ is  gap-ratio- and phase-independent and exactly coincides with the transmission-independent single-channel result $\kappa_\text{SNS}/\kappa_\text{N}$ at $\varphi=0$ and equal gaps, see Fig.~\ref{fig:singlechannel}. }
	\label{fig:average}
\end{figure}

In a diffusive conductor, the value of the transmission probability $D_n$ of the large number $\mathcal{N}$ of transmission eigenchannels has been found to be statistically distributed~\cite{Dorokhov1984Aug,Mello1989Sep} by the Dorokhov distribution $\rho(D_n)$. The average heat conductance is thus given by 
\begin{eqnarray}
\langle\kappa_\text{SNS}\rangle & = & \int_{0}^{1}dD_n \kappa_\text{SNS}^{(n)} \rho(D_n),\\
&&\text{with}\ \ \ \ \rho(D_n)=\mathcal{N}\ell\left(2 \mathcal{L} D_n \sqrt{1-D_n}\right)^{-1}
\end{eqnarray}
and with the single-channel contribution $\kappa_\text{SNS}^{(n)}$. Importantly, the transmission average $\int_{0}^{1}dD_n \mathcal{D}_n^e(E) \rho(D_n)$ is energy- and phase-independent~\cite{Virtanen2015Feb} and simply equals $\mathcal{N}\ell/\mathcal{L}$. Therefore, we can give the result for the average heat conductance as
\begin{eqnarray}
\label{eq_kappa_av}
\frac{\langle\kappa_\text{SNS}\rangle}{\langle\kappa_\text{N}\rangle} & = & 2-\frac{6}{\pi^2}\left[\left(\frac{|\Delta|}{k_\text{B}T}\right)^2\left(1-f(|\Delta|)\right)\right.\nonumber\\
& & \left.+2\frac{|\Delta|}{k_\text{B}T}\ln f(|\Delta|)-2\text{Li}_2\left(-e^{|\Delta|/k_\text{B}T}\right)\right],
\end{eqnarray}
with $\langle\kappa_\text{N}\rangle=\kappa_0\frac{\mathcal{N}\ell}{\mathcal{L}}$ and the dilogarithmic function $\text{Li}_2$. This shows that $\langle\kappa_\text{SNS}\rangle$ in a diffusive SNS junction is fully phase-independent and does not depend on the two gaps any longer, despite the junction being fully phase coherent (see also related discussions for the average charge conductance of normal, phase-coherent but diffusive conductors, for example in Refs.~\onlinecite{Nazarov1994Jul,Nazarov2009May}). The average heat conductance $\langle\kappa_\text{SNS}\rangle/\langle\kappa_\text{N}\rangle$ equals the single-channel result at vanishing phase difference $\varphi=0$ and equal gaps (where $\mathcal{D}_n^e=D_n$). This is shown  in   Fig.~\ref{fig:average}. The properties of the superconductors only enter Eq.~(\ref{eq_kappa_av}) via the magnitude of the (larger) gap $|\Delta|$ as a function of temperature~\cite{Virtanen2015Feb}. The temperature-dependent behavior of $\langle\kappa_\text{SNS}\rangle$, displayed in Fig.~\ref{fig:average}, reflects the exponential quasiparticle suppression at low temperatures and the gap closing at large temperatures close to $T_\text{crit}$. The normalconducting result is obtained at $T\geq T_\text{crit}$.
The averaging of effects due to the phase- and gap-differences can have severe consequences both for the field of phase-coherent caloritronics as well as for quasiparticle cooling with NS structures: the beneficial effect of the  phase- and gap-difference, which is exploited  in these research  fields, are fully suppressed when the junction is diffusive.

\section{Weak localization correction}\label{sec_weak}

 Due to enhanced backscattering of carriers~\cite{Altshuler1980Dec,Cahay1988Jun} and the resulting interference between time-reversed paths, the average quantum conductance is smaller than the classical one. This effect is known as weak localization. Equivalently to previous calculations for the (charge) conductance, the resulting correction to the average of the \textit{heat} conductance can be obtained considering the corresponding correction to the distribution of transmission eigenvalues $\delta\rho$,
\begin{align}
	\delta\kappa_\text{SNS}=\int_{0}^{\infty}dx \kappa_\text{SNS}^{(n)} \delta\rho(x).
\end{align}
Here, we use the parametrization $D_n=\frac{1}{\cosh^2x}$ for the transmission eigenvalues of the normal region, such that one can write $\delta\rho$  as~\cite{Beenakker1993Nov}
\begin{align}\label{Ddensity}
	\delta\rho(x)&=	\left(1-\frac{2}{\beta}\right)\Big[\frac{1}{4}\delta(x-0^+)+(4x^2+\pi^2)^{-1}\Big].
\end{align}
The parameter $\beta$ takes the values $\beta=1$ for time-reversal and spin-rotation symmetric systems (as the one considered here), hence we find $1-2/\beta=-1$. In the presence of a magnetic field, $\beta=2$ and the weak-localization correction vanishes. Since $\beta=4$ in the presence of strong spin-orbit coupling, one would in this case expect a relative factor $-1/2$ with respect to the weak-localization correction of the situation presented here. 
This correction to the transmission-probability distribution, and hence to the heat conductance, is of order $\mathcal{N}^0$ compared to the previously calculated $\langle\kappa_\text{SNS}\rangle$, which is of order $\mathcal{N}$. The weak localization correction is hence expected to be of importance in particular in devices with a rather small amount of channels, such as in recently developed hybrid superconductor-semiconductor devices~\cite{Larsen2015Sep,vanWoerkom2017Jun}. 
\begin{figure}[t]
			\includegraphics[scale=1]{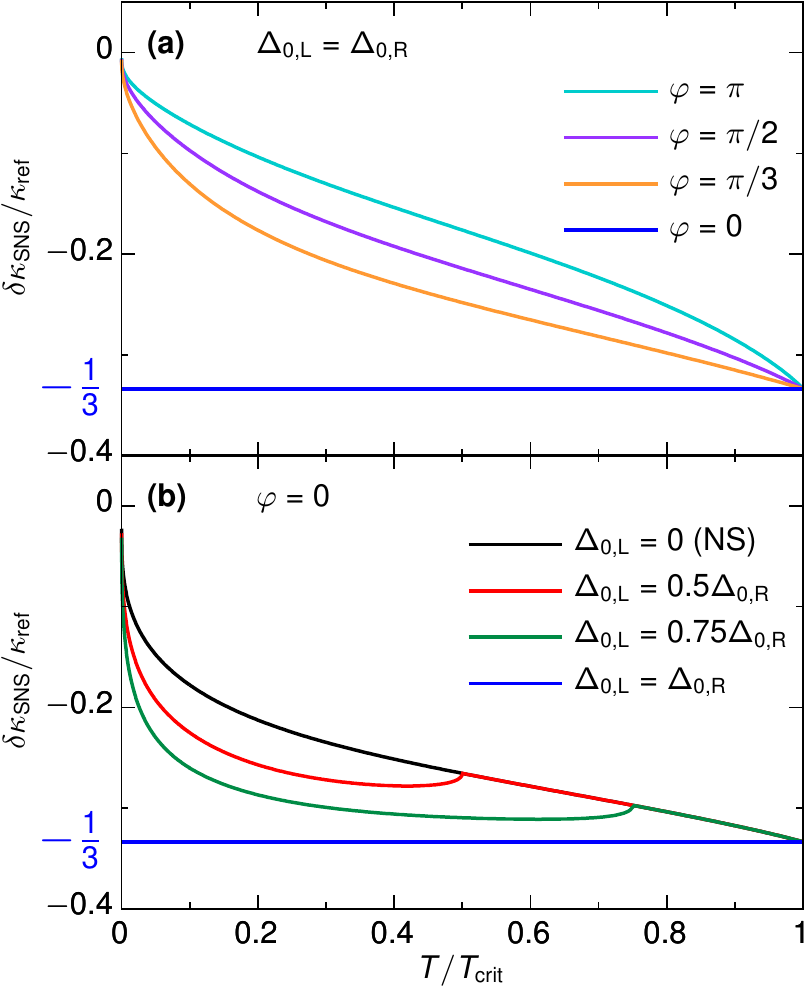}
	\caption{Weak localization correction to the heat conductance $\delta\kappa_\text{SNS}$ normalized with respect to the (length-independent) average heat conductance $\kappa_{\text{ref}}$ per channel, as function of the temperature. (a) Equal gaps $\Delta_{0,\text{L}}=\Delta_{0,\text{R}}$ at different phase differences $\varphi$ and (b)  different gap ratios $\Delta_{0,\text{L}}/\Delta_{0,\text{R}}$ at vanishing phase difference, $\varphi=0$.}
	\label{fig:WeakLoc}
\end{figure}

We show results for $\delta\kappa_\text{SNS}$ (with $\beta=1$) in Fig.~\ref{fig:WeakLoc}; we normalize with respect to 
\begin{equation}\label{eq_kref}
\kappa_{\text{ref}}:=\langle\kappa_\text{SNS}\rangle/(\mathcal{N}\ell/\mathcal{L}),
\end{equation}
which corresponds to the maximal average heat conductance per single channel. For equal gaps and vanishing phase difference, see blue lines in panel (a) and (b),  we recover the known result for the charge conductance $G$ of a normalconducting setup $\frac{\delta\kappa_\text{SNS}}{\kappa_{\text{ref}}}=\frac{\delta\langle G\rangle}{G_0}=-1/3$, with the conductance quantum $G_0=e^2/h$. The reason for this surprising occurrence of a generalized Wiedemann-Franz law in this limit, stating that the weak-localization correction to the heat conductance of the SNS junction $\frac{\delta\kappa_\text{SNS}}{\kappa_{\text{ref}}}$ and the weak-localization correction to the charge conductance of a normal conducting junction $\frac{\delta\langle G\rangle}{G_0}$ are equivalent, is the following: Eq.~(\ref{eq_fullTransmission}) for $\Delta_{0,\text{L}}=\Delta_{0,\text{R}}$ and $\varphi=0$ yields $\mathcal{D}_n^e=D_n$ for the total quasiparticle transmission above the gap. As a result the transmission of the normalconducting region, $D_n$, enters $\kappa_\text{SNS}$ in the same way as it enters $G$.

As soon as $\varphi\neq 0$ or $\Delta_{0,\text{L}}\neq\Delta_{0,\text{R}}$, the weak-localization correction to the heat conductance deviates from this value. This is in stark contrast to the previously presented channel average, where the effect of phase- and gap-difference is fully quenched. More specifically, the energy-dependence of the transmission amplitudes, induced by the presence of an Andreev bound state or by Andreev reflection at different gaps,  leads to a suppression of the weak localization effect, $\frac{|\delta\kappa_\text{SNS}|}{\kappa_{\text{ref}}}<\frac{1}{3}$. The suppression is strongest for low temperatures, where also the energy dependence of $\mathcal{D}_n^e(E)$, given in Eq.~(\ref{eq_fullTransmission}), is most pronounced. In particular in the zero-temperature limit $T/T_\text{crit}\rightarrow 0$, we find $\frac{\delta\kappa_\text{SNS}}{\kappa_{\text{ref}}}\rightarrow 0$.

\section{Heat conductance fluctuations}\label{sec_fluct}
\begin{figure}[t!]
	\includegraphics[scale=1]{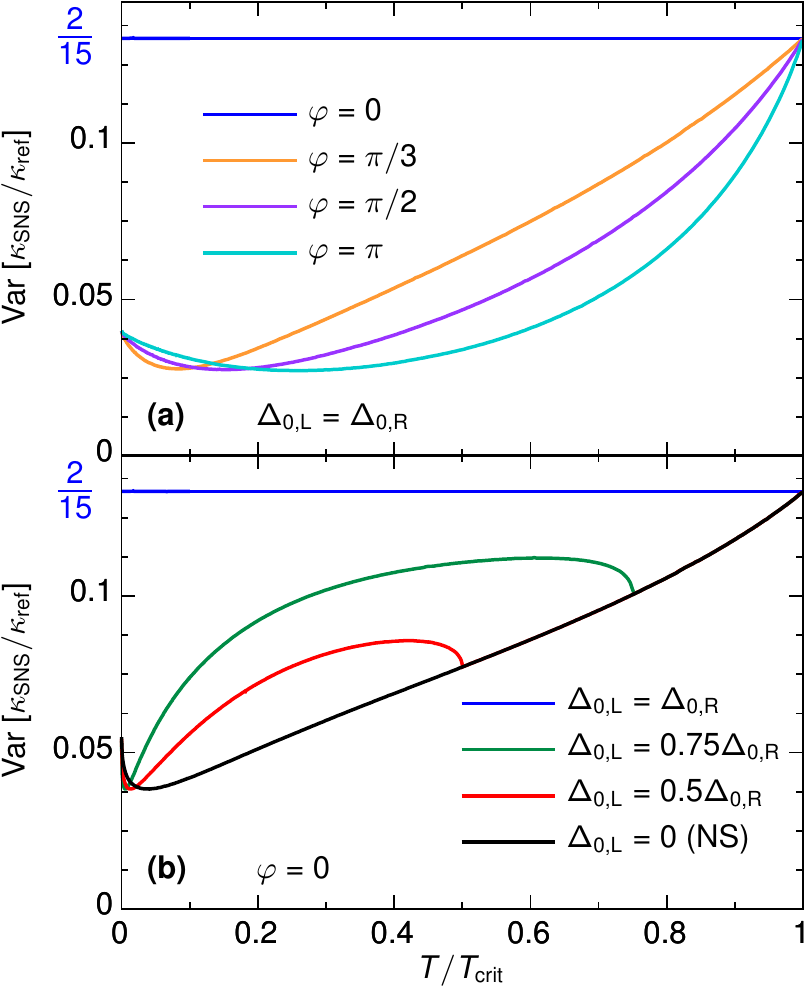}
	\caption{Heat conductance fluctuations  normalized with respect to the (length-independent) average heat conductance per channel $\text{Var}\left[\kappa_\text{SNS}/\kappa_\text{ref}\right]$, as function of the temperature. (a) Equal gaps $\Delta_{0,\text{L}}=\Delta_{0,\text{R}}$ at different phase differences $\varphi$ and (b)  different gap ratios $\Delta_{0,\text{L}}/\Delta_{0,\text{R}}$ at vanishing phase difference, $\varphi=0$. }
	\label{fig:Fluct}
\end{figure}
Finally, we want to address the variance of the heat conductance, which constitutes the heat transport analogue to the well-known charge conductance fluctuations~\cite{Lee1985Oct,Altshuler1985Jun,Imry1986Mar,Altshuler1991Jul}. Thanks to the eigenchannel decomposition of the full transmission matrix, leading to Eq.~(\ref{eq_fullTransmission}), the variance  $\text{Var} \left[\kappa_\text{SNS}\right]$ can directly be computed from~\cite{Beenakker1993Nov,Beenakker1997Jul}
\begin{align}
	\text{Var}\left[ \kappa_\text{SNS}\right]=\frac{1}{2\pi^2}\int_{0}^{\infty}dx \int_{0}^{\infty}dx^\prime\left( \frac{d\kappa^{(n)}_\text{SNS}(x)}{dx}\right) \\
	\times\left( \frac{d\kappa^{(n)}_\text{SNS}(x^\prime)}{dx^\prime}\right) \text{ln}\left( \frac{1+\pi^2(x-x^\prime)^{-2}}{1+\pi^2(x+x^\prime)^{-2}}\right), \nonumber
\end{align}
using the previously introduced parametrization of $D_n(x)$. Importantly, this term is of order $\mathcal{N}^0$; for the conductance $G$ of a normalconducting junction in the diffusive limit, the conductance fluctuations  take the \textit{universal} value  $\text{Var}\left[G/G_0\right]=2/15$ for the type of system we are considering here. In particular, via the Wiedemann-Franz law this directly entails that the variance of the heat conductance of the \textit{normalconducting} junction, $\kappa_\text{N}=\kappa_0\sum_{n=0}^\mathcal{N}D_n$,  is given by 
$\text{Var}\left[\kappa_\text{N}/\kappa_0\right]  =  \frac{2}{15}$. 
The variance of the normalconducting heat conductance, 
\begin{equation}
\text{Var}[\kappa_\text{N}]= \frac{2}{15}\left(\frac{\pi^2k_\text{B}^2}{3h}\right)^2T^2,
\end{equation}  is hence also universal ---up to a factor $T^2$, which is expected for the heat conductance (having a temperature-dependent heat-conductance quantum $\kappa_0$).

Due to the complex energy- and phase-dependence of the transmission probability $\mathcal{D}_n^e$, the fluctuations of the heat conductance of the diffusive SNS junction $\text{Var}\left[\kappa_\text{SNS}\right]$ are more intricate ---only in the limit of equal gap and vanishing phase difference does the variance equal $\text{Var}\left[\kappa_\text{SNS}/\kappa_{\text{ref}}\right]=\frac{2}{15}$. Namely, we find that the previously discussed generalization of the Wiedemann-Franz law applies to the (heat) conductance fluctuations as well.

In Fig.~\ref{fig:Fluct}, we show $\text{Var}\left[\kappa_\text{SNS}/\kappa_{\text{ref}}\right]$ as a function of temperature, both for equal gaps $\Delta_{0,\text{L}}=\Delta_{0,\text{R}}$ at different phases $\varphi$, as well as for $\varphi=0$ but different gap ratios $\Delta_{0,\text{L}}/\Delta_{0,\text{R}}$. The blue lines show the universal result in the equal-gap, zero phase limit. Both panels demonstrate that the heat conductance fluctuations are sensitive to the gap difference as well as to the phase-difference between the two contacts. Importantly, despite these nontrivial dependencies leading to a suppression with respect to $2/15$, the heat conductance fluctuations stay however close to universal. The difference of fluctuation values is less than an order of magnitude at its biggest discrepancy. In particular, even in the zero-temperature limit, the fluctuations of the heat conductance, $\text{Var}\left[\kappa_\text{SNS}/\kappa_{\text{ref}}\right]$, approach a finite, gap-dependent value.

\section{Conclusions} 

We have shown an analysis of the phase- and gap-dependent heat conductance of SNS and NS hybrid junctions.
In particular, this work demonstrates how the phase- and gap-dependent heat conductance of a hybrid superconducting junction is impacted by the properties of a diffusive junction. While the value of the heat conductance in the diffusive limit leads to a complete quenching of effects induced by the gap- and phase-difference of the superconducting contacts, the dependence on these parameters persists in weak-localization corrections and conductance fluctuations. We find that despite its intricate phase- and gap-dependence, the heat conductance fluctuations stay close to universal, similar to the famous charge-conductance counterpart.

We expect that experimental verifications of the predicted phenomena are possible for diffusive metallic~\cite{Dubos2001Jan} and semiconducting devices with statistically distributed channel transmissions~\cite{vanWoerkom2017Jun}, exploiting advanced thermometry as recently used for caloritronics measurements, e.g., in Ref.~\onlinecite{Giazotto2012Dec} and further developed in Refs.~\onlinecite{Wang2018Jan,Karimi2018Nov}. 

\acknowledgments 
We thank A. Geresdi, T. L\"ofwander, and H. Pothier for interesting and helpful discussions, and T. L\"ofwander and K. Seija for support with the self-consistent gap data. This work has been financially supported by the Knut and Alice Wallenberg foundation via a KAW Fellowship (J. Splettstoesser), the Swedish Vetenskapsr\aa det (J. Splettstoesser and F. Hajiloo), the COST action MP1209 (F. Hajiloo) and the Deutsche Forschungsgemeinschaft 353 (DFG) under Grant No. CA 1690/1-1 (F. Hassler).


%

\end{document}